\documentclass[prb,twocolumn,showpacs]{revtex4}

\usepackage{rotating}
\usepackage{graphicx}
\usepackage{amssymb}

\newcommand{\la}{\mathord{\langle}}
\newcommand{\ra}{\mathord{\rangle}}

\begin{document}

\title
{Anisotropic susceptibility of ferromagnetic ultrathin Co films on vicinal Cu}

\author
{P. J. Jensen}
\email
[Corresponding author. Electronic adress: ]{jensen@physik.fu-berlin.de}
\affiliation
{Institut f\"ur Theoretische Physik, Freie Universit\"at 
Berlin, Arnimallee 14, D-14195 Berlin, Germany}

\author
{S. Knappmann}
\altaffiliation
{present adress: Corporate Research, Thomson Multimedia, DTB, 
Hermann-Schwer-Str.3, D-78048 Villingen-Schwenningen, Germany}

\author
{W. Wulfhekel}
\altaffiliation
{present adress: Max-Planck-Institut f\"ur 
Mikrostrukturphysik, Weinberg 2, D-06120 Halle, Germany}
\affiliation
{Institut f\"ur Grenz- und Oberfl\"achenphysik (ISG3),
Forschungszentrum J\"ulich, D-52425 J\"ulich, Germany}

\author
{H. P. Oepen}
\email
{oepen@physnet.uni-hamburg.de}
\affiliation
{Institut f\"ur Angewandte Physik, Universit\"at Hamburg, 
Jungiusstr.11, D-20355 Hamburg, Germany}

\date{\today}
 
\begin{abstract} 
We measure the magnetic susceptibility of ultrathin
Co films with an in-plane uniaxial magnetic anisotropy grown on 
a vicinal Cu substrate. 
Above the Curie temperature the influence of the magnetic anisotropy 
can be investigated by means of the parallel and transverse  
susceptibilities along the easy and hard axes. 
By comparison with a theoretical analysis of the susceptibilities 
we determine the isotropic exchange interaction and the magnetic 
anisotropy. These calculations are performed in the 
framework of a Heisenberg model by means of a many-body Green's function 
method, since collective magnetic excitations are  
very important in two-dimensional magnets. 
\end{abstract}
\maketitle                                           

\section{Introduction}
The  investigation   of  the  magnetic   properties  of  ferromagnetic
ultrathin films  is a field of intense  current interest.\cite{HeB94}
Among the different experimental methods the measurement of 
the magnetic susceptibility $\chi(T)$ is a very 
powerful method for the analysis of such thin film 
systems.\cite{WKG94} The singularity of $\chi(T)$ corresponds to the 
onset of a ferromagnetic state, i.e.\ to the occurrence of a nonvanishing 
magnetization $m(T)=|\mathbf{m}(T)|$ for temperatures below the 
(ferromagnetic) Curie temperature $T_C$. For $T\gg T_C$ the 
\textit{inverse susceptibility} $\chi^{-1}(T)$
exhibits the linear (Curie-Weiss-) behavior: 
$\chi^{-1}(T)\propto T-T_C^\mathrm{para}$.  
The paramagnetic Curie temperature $T_C^\mathrm{para}\ge T_C$ is 
obtained from the extrapolation of this linear behavior to 
$\chi^{-1}(T)=0$, which corresponds to the Curie 
temperature calculated in the mean field approximation.\cite{Tya67} 
For an isotropic ferromagnet the behavior of $\chi(T)$ does not 
depend on the lattice orientation. 

In the collectively ordered magnetic state the direction of the 
magnetization is determined by magnetic anisotropies, which 
are the \textit{free energy differences} between the hard and 
easy magnetic directions. Due to their relativistic origin 
resulting from the spin-orbit interaction they are 
usually much smaller than the isotropic exchange.
As obtained in experiments the anisotropies depend on 
temperature and are expected to vanish above the Curie 
temperature.\cite{Cal66} It is known 
from general considerations\cite{PT77} 
that the mentioned singularity (or maximum) of the susceptibility 
is only observed if $\chi(T)$ is measured along 
\textit{easy} magnetic directions. Corresponding experiments have 
been performed for bulk systems,\cite{SuW70} a thin film investigation
has been reported for the Fe/W(110) system.\cite{BAP94} 
Thus, a signature of the anisotropy is also present in the
paramagnetic state above $T_C$. 

At first we comment on the fact that the anisotropy is noticeable also 
for $T>T_C$. We like to stress the fact that the \textit{microscopic} 
anisotropy, e.g.\ the single-ion uniaxial anisotropy $K_2$ 
as present in a Heisenberg Hamiltonian, is different from the 
\textit{effective}, temperature dependent anisotropy $\mathcal{K}_2(T)$ 
as measured for a collectively ordered magnetic state.\cite{Cal66}  
The effective anisotropy is equal to the microscopic one for $T=0$, 
thus $\mathcal{K}_2(0)=K_2$. When treated as a 
\textit{perturbation} to the exchange interaction, $\mathcal{K}_2(T)$ 
indeed vanishes at $T_C$.\cite{Cal66} However, 
a vanishing  effective  anisotropy  for  $T>T_C$  does  not
indicate that the microscopic anisotropy vanishes either, or that the 
underlying spin-orbit coupling is strongly varying with 
temperature. A noticeable drop of the spin-orbit coupling 
is expected on a larger temperature scale.\cite{MHB96} 
Thus, a single magnetic moment in the paramagnetic state 
is still subject to the anisotropy even if the net magnetization is
zero. Here a free energy  difference 
between the easy  and hard  magnetic  directions is also present 
('paramagnetic  anisotropy'), exhibiting a temperature behavior as
$\propto(K_2)^2/k_BT$ for  $K_2\ll k_BT$, with $k_B$ the Boltzmann 
constant.\cite{Jenup} Evidently, the paramagnetic 
anisotropy is rather small as $K_2$ is small compared  to the 
exchange interaction  $J\propto k_BT_C$. 

In the present study we will show that the anisotropy -- although small 
-- has a sizable effect on the susceptibilities in the paramagnetic 
state of ultrathin films in particular when approaching the Curie temperature. 
Whereas a vast amount of susceptibility data are available 
for various systems,\cite{WKG94} to our knowledge 
the different behavior of $\chi(T)$ measured along the easy and 
hard magnetic directions has not been exploited to gain information 
about thin films. In this paper we 
report measurements of $\chi(T)$ for ferromagnetic ultrathin Co film 
grown on a vicinal Cu substrate. This system exhibits an \textit{in-plane} 
two-fold (uniaxial) magnetic anisotropy due to the presence of regularly 
distributed steps in the Cu surface, with the easy axis directed along 
the steps.\cite{BLO92} 
We find strong differences for the magnetic susceptibilities along the 
easy ($z$-) and hard ($x$-) in-plane magnetic directions. 
With the help of an anisotropic Heisenberg model solved within a 
many-body Green's function method 
we are able to perform a quantitative comparison with experiments.  
Furthermore, we demonstrate how the exchange interaction and magnetic 
anisotropy can be extracted from these susceptibilities. 
In the previous study on Fe/W(100) films only a qualitative 
comparison in the framework of a renormalization treatment was 
possible.\cite{BAP94} Our present approach represents a new method to 
study quantitatively magnetic properties of ultrathin
ferromagnetic films with the help of high-accuracy susceptibility 
measurements above $T_C$. 

The paper is organized as follows. In Secs.II and III we describe the
experimental methods and the theoretical model. Results from 
measurements and calculations are presented in Sec.IV. 
A discussion and conclusion is given in Sec.V. 

\section{Experimental aspects}
The experiments, including film preparation and investigation of the 
magnetic properties, were performed under Ultra High Vacuum (UHV) 
conditions (base pressure $\sim 10^{-10}$ Torr) in the same chamber. The 
film characterization and surface preparation were made via Auger 
Electron Spectroscopy (AES), Low Energy Electron Diffraction (LEED) and 
Medium Energy Electron Diffraction (MEED). 

Vicinal Cu(11n) surfaces with $n=17$ are used as templates for the 
ultrathin Co 
films. These surfaces have been well studied by means of helium 
scattering\cite{LPK83} and Scanning Tunneling Microscopy (STM).\cite{FGP91}
Microscopically, the surfaces consist of terraces 
with the normal oriented along the (001) direction, and an average 
terrace width of $n/2$ atomic distances. The terraces are 
separated by mono-atomic steps which are aligned along the [$1\bar10$] 
in-plane direction. Step bunching has not been observed. 
The substrate was cleaned and prepared by cycles of Ar sputtering (600~eV) 
and subsequent annealing ($T>670{}^\circ$C). The quality 
of the surface structure was confirmed via LEED and MEED. Pronounced 
splitting of regular lattice spots were found, indicating the periodic 
step arrangement on the surface. 

The Co films were grown at $T=45{}^\circ$C with a rate of 
one monolayer (ML) per minute. During electron beam evaporation the 
pressure did not exceed $5\cdot 10^{-10}$ Torr.  The growth 
process was monitored by measuring MEED intensity 
oscillations, which were used for thickness calibration.

The magnetic characterization has been performed \textit{in situ} using 
the longitudinal magneto-optic Kerr effect 
(MOKE).\cite{BKO94,OKW95,WKO96} Hysteresis loops were obtained in static 
magnetic fields up to $B\sim30$ mT. With the same optical setup the 
magnetic susceptibility $\chi(T)$ was studied. For that purpose, 
the change of the Kerr ellipticity in an applied \textit{ac} magnetic 
field has been determined via 
phase sensitive detection. No additional bias fields were used, hence 
the zero field \textit{ac} susceptibility is measured.\cite{BKO94} 
In this paper we monitor the static susceptibility, which is 
obtained for sufficiently low magnetic fields and frequencies. 
The modulation frequency was set to approximately 110 Hz, while the 
modulation amplitude $B_\mathrm{mod}$ has been varied on purpose. 
$\chi(T)$ was measured for different in-plane directions 
of the applied magnetic field, with angular uncertainties of about
$\pm\,5\,{}^\circ$. Due to the optical setup we have monitored 
the magnetic response 
along the magnetic field direction (longitudinal susceptibility). 

Sample heating is a very delicate issue while performing zero-field 
measurements. At first the heating has to be performed quasi-statically 
to achieve an equilibrium phase transition. Secondly, any spurious 
magnetic fields have to be prevented as they will alter the 
manifestation of the phase transition, particularly when investigating 
the zero field susceptibility. For the latter reason we have used an 
external light source for heating the front side of the sample. Due to 
this arrangement some scattering of our data appear as the film 
warms up quicker than the thermocouple, located at the rear side of
the sample, on every change of the radiation intensity. This affects the 
measurements and causes some of the apparently periodic modulations 
in the plots for $\chi^{-1}(T)$. 

\section{Theoretical model}
To calculate the susceptibility of the ferromagnetic ultrathin film 
with an in-plane uniaxial anisotropy we apply an anisotropic 
Heisenberg Hamiltonian with localized three-component spins 
$\mathbf{S}_i$ with spin quantum number $S$ on lattice sites $i$: 
\begin{equation} \mathcal{H} =-\frac{1}{2}\;\sum_{i,j} 
\Big( J_{ij}\;\mathbf{S}_i\,\mathbf{S}_j + D_{ij}\;S_i^z\,S_j^z\Big) 
-g\,\mu_B\sum_i \mathbf{B}\;\mathbf{S}_i  \;. \label{e1} 
\end{equation}
A thin film with $L$ atomic layers is assumed, spanned by the $xz$- plane.   
$J_{ij}$ is the isotropic exchange interaction between spins $i$ and $j$. 
The last term is the Zeeman energy, with the magnetic field 
${\bf B}=(B_x,0,B_z)$ confined to the film plane, where 
$g$ is the Land\'e factor and $\mu_B$ the Bohr magneton. 
The uniaxial magnetic anisotropy within a Heisenberg Hamiltonian is 
usually represented by a \textit{magneto-crystalline single-ion anisotropy}, 
$-\sum_i K_{2,i}\,(S_i^z)^2$, impling a spin quantum number $S\ge1$.  
However, such a single-ion anisotropy complicates considerably the 
solution with the method applied in this study.\cite{FJK00} 
Thus, for simplicity we consider an \textit{exchange anisotropy} 
$-(1/2)\sum D_{ij}\,S_i^z\,S_j^z$ between nearest neighbor spins. 
Although originating from very different physical mechanisms, 
the anisotropic properties obtained from a single-ion
term and an exchange anisotropy are quite similar, if one assumes 
$K_2\sim(q/2)\,D$, with $D\equiv D_{ij}$ and $q$ the coordination 
number.\cite{FJK00,Lin71} A positive value for $D$ indicates the 
easy direction to be parallel to the $z$- axis. Note that
ferromagnetic thin films with a strong surface anisotropy sometimes 
exhibit a magnetization perpendicular to the film plane. The 
interpretation of the susceptibility of such a system is more
complicated due to the shape anisotropy resulting from the dipole 
interaction. The magnetic ground state in this case is a stripe-domain 
structure, and not the single-domain state.\cite{YaG88} This complication
vanishes for an \textit{in-plane} uniaxial anisotropy, since then the 
dipole interaction favors a single-domain, ferromagnetic ground state.  
Thus, this coupling is not considered explicitely in the present study. 

The Hamiltonian Eq.(\ref{e1}) is solved by a many-body Green's function 
approach,\cite{Tya67} which is  suited to consider simultaneously both 
expectation values $m_{z,i}(T)=\la S_i^z\ra$ and 
$m_{x,i}(T)=\la S_i^x\ra$.\cite{Jenup} 
Furthermore, collective magnetic excitations (spin waves) are taken 
into account, which are particularly important for low-dimensional 
systems. The long range magnetic order of an isotropic two-dimensional  
Heisenberg magnet becomes   unstable  against   collective   magnetic
excitations with long wavelengths (Mermin-Wagner-theorem).\cite{Mer66}
Already weak anisotropies, however, induce a magnetization with a 
Curie  temperature of the order of the exchange coupling.\cite{Kit51} 

We consider the set of anticommutator Green's functions in frequency 
space, $G_{ij}^\alpha(\omega)=\la\la S_i^\alpha;S_j^-\ra\ra_\omega$, 
where the ladder operators $S^\pm=S^x\pm iS^y$ have been introduced, 
and $\alpha=+,-,z$. These Green's functions are  solved in the
usual  way by  the equation  of motion.\cite{Tya67} 
The vanishing eigenvalues occurring in the set of equations  
motivate the use of \textit{anticommutator} Green's functions. The
higher-order Green's functions appearing within this procedure are 
approximated   by   the 
generalized Tyablikov-decoupling (RPA) for $i\neq k$,\cite{Tya67,FJK00}
\begin{equation} \la\la S_k^\alpha\,S_i^\beta;S_j^-\ra\ra\simeq
\la S_k^\alpha\ra\la\la  S_i^\beta;S_j^- \ra\ra+
\la S_i^\beta\ra\la\la  S_k^\alpha;S_j^- \ra\ra \,. \label{e2} 
\end{equation}  
Interactions between spin waves are partly taken into account, 
allowing for the determination of the magnetic properties with a 
reasonable accuracy in the whole temperature range. 
It has been shown recently that the magnetization and the Curie 
temperature of a weakly anisotropic $S=1$- Heisenberg monolayer 
calculated by this approach agrees very well with the values 
as obtained from a Quantum Monte Carlo method.\cite{HFK02} 

The set of linear equations of the corresponding correlation functions 
$\la S_j^-S_i^\alpha\ra$ can be solved numerically for films with an 
arbitrary number $L$ of layers, for inhomogeneous coupling constants 
$J_{ij}$ and $D_{ij}$, and for arbitrary spin quantum numbers $S$. 
This will be investigated in a forthcoming study.\cite{Jenup} 
In the remainder of this section we present two cases 
for which analytical solutions can be derived. 

First, to give some insight in the structure of the solutions we 
consider a homogeneous square (001) ferromagnetic monolayer 
($L=1$ and $q=q_\|=4$) with spins $S=1/2$.  The 
coupling constants are put equal to $J_{ij}=J>0$ and $D_{ij}=D>0$ if 
$i$ and $j$ are nearest neighbors, and zero otherwise. 
A Fourier transformation into the two-dimensional wave vector space  
$\mathbf{k}\equiv\mathbf{k}_\|=(k_x,k_z)$ is applied.  
By considering the properties for $S=1/2$- spin operators, the 
magnetization components $m_x(T)$ and $m_z(T)$ are given implicitely 
by the equations 
\begin{eqnarray} 
\frac{1}{2} &=& m_z \; \frac{1}{N} \sum_\mathbf{k} \frac{H}{H_z}\, 
\coth(H/2\,k_BT) \,, \label{e3a} \\ 
 \frac{1}{2} &=& m_x \; \frac{1}{N} \sum_\mathbf{k} \frac{H}{\tilde H_x}\, 
\coth(H/2\,k_BT) \,, \label{e3b} \end{eqnarray}
with the denotations 
\begin{eqnarray} H &=& \sqrt{H_z^2+H_x\,\tilde H_x} \,, \label{e4} \\
H_z &=& m_z\,[J\,(q_\|-\gamma_\mathbf{k})+D\,q_\|]+g\,\mu_B\,B_z \,,
\label{e5} \\ 
\tilde H_x &=& m_x\,J\,(q_\|-\gamma_\mathbf{k})+g\,\mu_B\,B_x \,, 
\label{e6} \\ 
H_x &=& m_x\,[J\,(q_\|-\gamma_\mathbf{k})-D\,\gamma_\mathbf{k}]
+g\,\mu_B\,B_x \,, \label{e7} \end{eqnarray}
where $\gamma_\mathbf{k}=2\,[\cos(k_x/a_0)+\cos(k_z/a_0)]$, $a_0$ 
the lattice constant, and $N$ the number of \textbf{k}-points in the 
first Brillouin zone. From Eqs.(\ref{e3a}) and (\ref{e3b}) the 
susceptibilities $\chi_{zz}$ 
and $\chi_{xx}$ along the easy and hard axes, which we denote by 
'parallel' and 'transverse' susceptibilities, 
will be determined numerically. The Curie temperature 
$T_C\equiv T_C^\mathrm{RPA}$ is calculated from 
\begin{equation} 
\frac{1}{4\,k_BT_C} = \frac{1}{N} \sum_\mathbf{k} 
\Big(J\,(q_\|-\gamma_\mathbf{k})+D\,q_\|\Big)^{-1} 
\,. \label{e8} \end{equation}
Note that the value of $T_C$ is determined 
not only by the isotropic exchange interaction $J$ but depends also 
on the exchange anisotropy $D$.\cite{FJK00,Lin71,Kit51}
The mean field approximation (MFA) is obtained by putting
$\gamma_\mathbf{k}=0$ in Eqs.(\ref{e5}) -- (\ref{e8}), yielding the 
corresponding ordering temperature $4\,k_BT_C^\mathrm{MFA}=q_\|\,(J+D)$. 
We point out that the paramagnetic Curie temperature 
$T_C^\mathrm{para}$ calculated within the Green's function method is 
equal to $T_C^\mathrm{MFA}$.\cite{Tya67} 

Secondly, to allow for a quantitative comparison with the 2~ML Co/Cu 
thin film system as investigated experimentally in the present study, we 
consider a homogeneous fcc (001) film with $L=2$ layers and spins $S=1/2$. 
For a magnetic field along the easy ($z$-) axis ($B_x=m_x(T)=0$) 
we calculate the magnetization component $m_z(T)$ from 
\begin{equation} 
\frac{1}{2} = m_z \; \frac{1}{N} \sum_\mathbf{k} 
\frac{\sinh(\bar H_\|/k_BT)}{\cosh(\bar H_\|/k_BT)
-\cosh(m_z\,H_\perp/k_BT)} \,, \label{e9}  \end{equation} 
denoting 
\begin{eqnarray} 
\bar H_\| &=&  m_z\,H_\|+g\,\mu_B\,B_z \,, \label{e10} \\
H_\| &=& J\,(q_\|+q_\perp-\gamma_\mathbf{k})+D\,(q_\|+q_\perp) \,, 
\label{e11} \\ 
H_\perp &=& J\,\lambda_\mathbf{k} \,, \label{e12} \end{eqnarray}
where $q_\perp=4$ is the coordination number between neighboring layers 
and $\lambda_\mathbf{k}=4\,\cos(k_x/2\,a_0)\,\cos(k_z/2\,a_0)$. For 
comparison a simple-cubic (001) film is characterized by 
$q_\perp=\lambda_\mathbf{k}=1$. 
The Curie temperature for this two-layer film is given by 
\begin{eqnarray} 
\frac{1}{4\,k_BT_C} &=& \frac{1}{N} \sum_\mathbf{k} 
\frac{H_\|}{(H_\|)^2-(H_\perp)^2} \,. \label{e12a} 
\end{eqnarray}  

From a fit to experimental data the coupling constants are 
determined. This can be done by using both $\chi_{zz}(T)$ 
and $\chi_{xx}(T)$, see 
Eqs.(\ref{e3a}) and (\ref{e3b}). Alternatively, on can employ 
solely $\chi_{zz}(T)$, since the increase of $\chi_{zz}^{-1}(T)$ 
for $T\gtrsim T_C$ depends sensitively on the anisotropy. The 
latter method is used for the determination of $J$ and $D$ of the 
present 2~ML case, since for 2~ML an analytical 
expression is only available for $m_z(T)$, see Eq.(\ref{e9}). 
A corresponding expression for $m_x(T)$ along
the hard-axis needs for additional numerical work.
\begin{figure}[t] \label{fig1} 
\hspace*{-0.5cm} \includegraphics[width=6cm,angle=-90,bb=50 100 520 750,clip]
{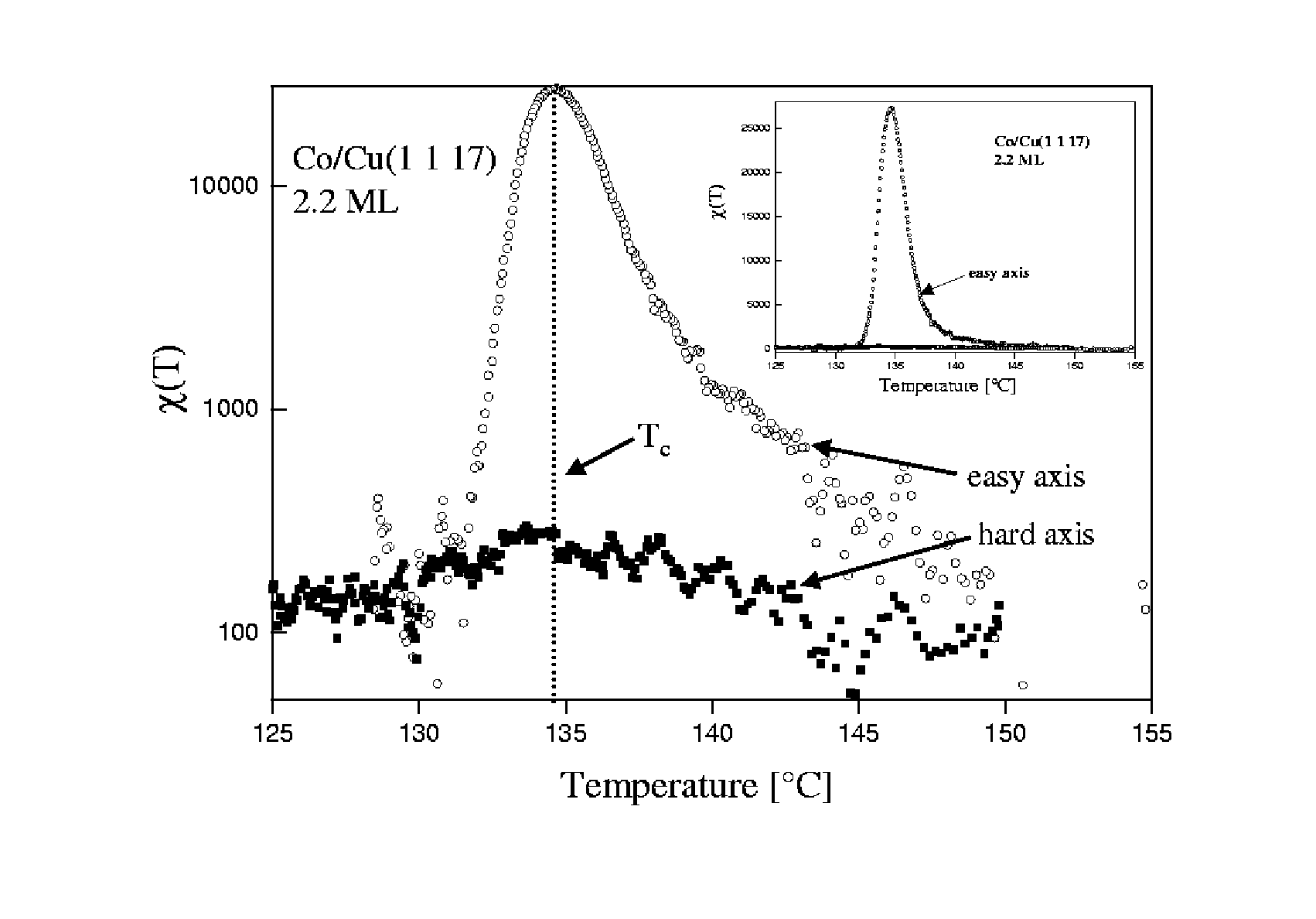}
\caption{ Parallel and transverse susceptibilities $\chi_{zz}(T)$ 
(open symbols) and $\chi_{xx}(T)$ (closed symbols) in SI-units
measured along the easy and hard in-plane directions 
of a vicinal Co/Cu(1~1~17) thin film as function of temperature. 
The nominal film thickness is 2.2~ML, the Curie temperature is
$T_C=134.7{}^\circ$C. The modulation
amplitude is $B_\mathrm{mod}=4.86\;\mu$T along the easy axis, and 
$B_\mathrm{mod}=24.3\;\mu$T along the hard axis, respectively. To
present $\chi(T)$ in SI units we have assumed that the saturation 
magnetization corresponds to the bulk Co value
$m_s=1.43\cdot10^6$~A/m. A half-logarithmic plot has been used in order  
to show the temperature behavior of both susceptibilities 
simultaneously. In the inset the corresponding linear plot is shown. 
}\end{figure} 

\section{Results}      
The transition temperatures for the Co/Cu(001) and the Co/Cu(1 1 17) 
thin films exhibit a similar dependence on film 
thickness.\cite{Kna94} 
Due to the high instability against surface diffusion
of Co films on Cu,\cite{SAI93,GSI95} and the steep increase of $T_C$
with increasing film thickness,\cite{SBS90} a very thin Co film of 
about 2.2~ML was 
chosen with $T_C$ well below $180{}^\circ$C.\cite{SAI93} 
The real part of the susceptibility $\chi(T)$ in SI units 
as a function of the temperature is shown in Fig.1. Both parallel and 
transverse susceptibilities for magnetic field directions along the 
easy and hard axes are displayed. The semi-logarithmic plot allows for
a comparison of both quantities. The inset shows the corresponding 
linear plot. The parallel susceptibility exhibits a strong peak at 
$T_C=134.7{}^\circ\mathrm{C}=407.8$~K, with a full width half maximum 
(FWHM) of 2.7~K for the actual modulation amplitude 
$B_\mathrm{mod}=4.86\;\mu$T. The FWHM can be reduced to values 
around 1.5~K for smaller modulation amplitudes.\cite{OKW95}  
For $B_\mathrm{mod}\gtrsim 1.62\;\mu$T an imaginary part 
of the parallel susceptibility is observed, while for
the smallest applied modulation field ($B_\mathrm{mod}=0.81\;\mu$T) 
the imaginary part vanishes.\cite{Kna94}

While a peak was found in the susceptibility $\chi_{zz}$ along the 
easy axis, the magnetic response $\chi_{xx}$ along the hard 
axis was not detectable for small $B_\mathrm{mod}\lesssim5\;\mu$T. 
In order to increase the magneto-optical signal and thus 
the detection limit, the modulation amplitude along the hard axis was 
increased by a factor of five. However, the response is still small, 
exhibiting a weak maximum near $T_C$, as shown in Fig.1. 
Obviously, the phase transition is reflected only weakly 
in the transverse susceptibility signal. Thus, the phase transition 
for this thin film system with an in-plane magnetization exhibits 
a strongly anisotropic behavior. 
\begin{figure}[t] \label{fig2} 
\hspace*{-0.5cm} \includegraphics[width=6cm,angle=-90,bb=90 120 520 720,clip]
{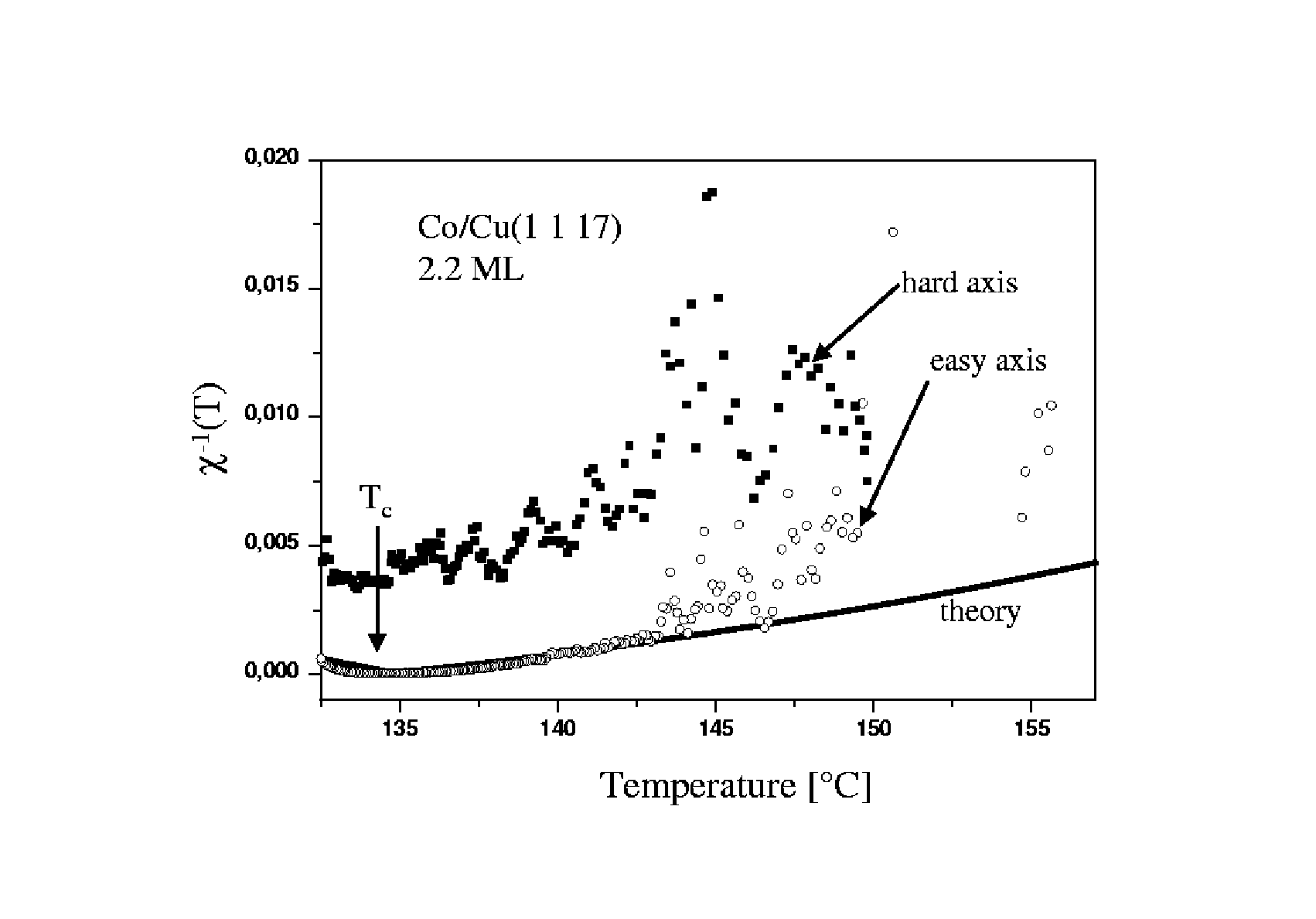}
\caption{
Inverse susceptibilities $\chi_{zz}^{-1}(T)$ and 
$\chi_{xx}^{-1}(T)$ in SI-units 
near the Curie temperature $T_C=134.7{}^\circ$C. The 
data are taken from Fig.1. In addition the full line shows 
$\chi_{zz}^{-1}(T)$ as calculated from Eq.(\ref{e9}), using the isotropic 
exchange coupling $J=44.1$~meV/bond and the exchange anisotropy 
$D=7.0$~mK/atom. 
}\end{figure} 

Fig.2 displays the temperature dependence of the \textit{inverse} 
parallel and transverse susceptibilities $\chi_{zz}^{-1}(T)$ and 
$\chi_{xx}^{-1}(T)$ around the Curie temperature. For elevated 
temperatures $T\gtrsim 145{}^\circ$C the scattering of the data
points is strong, and the temperature dependence 
of $\chi_{zz}^{-1}(T)$ and $\chi_{xx}^{-1}(T)$ cannot be given
precisely. As mentioned in Sec.II this scattering is due to the 
fact that the signal becomes very small, and that 
the measuring time cannot be increased appropriately as the system 
properties might change due to the onset of surface diffusion. 
The oscillations obtained for $\chi_{xx}^{-1}(T)$ are most likely 
caused by too large steps of changes of the heating power, 
which were indeed larger than in case of $\chi_{zz}^{-1}(T)$. 
We have not systematically explored these effects. 
On the other hand, within the temperature range between 
$T=T_C$ and $T=150{}^\circ$C the behavior of $\chi_{zz}^{-1}(T)$ 
and $\chi_{xx}^{-1}(T)$ is clearly resolved. We remark that 
the inverse susceptibilities cannot be described by straight
lines in this temperature range, rather they exhibit an upward
curvature. Evidently, $\chi_{xx}^{-1}(T)$ 
is shifted upwards with respect to $\chi_{zz}^{-1}(T)$ by an almost 
constant, temperature independent 
amount. Thus, at any temperature above $T_C$ the inverse 
susceptibility $\chi_{xx}^{-1}(T)$ along the hard axis is larger 
than the corresponding value $\chi_{zz}^{-1}(T)$ along the easy axis. 
This is in accordance with measurements for bulk magnets,\cite{SuW70} 
and is also expected theoretically.\cite{PT77} 
We will show that the temperature range as displayed in Fig.2
is still far below the linear regime of the inverse susceptibilities. 
\begin{figure}[t] \label{fig3} 
\hspace*{-0.5cm} \includegraphics[width=6cm,angle=-90,bb=40 30 580 750,clip]
{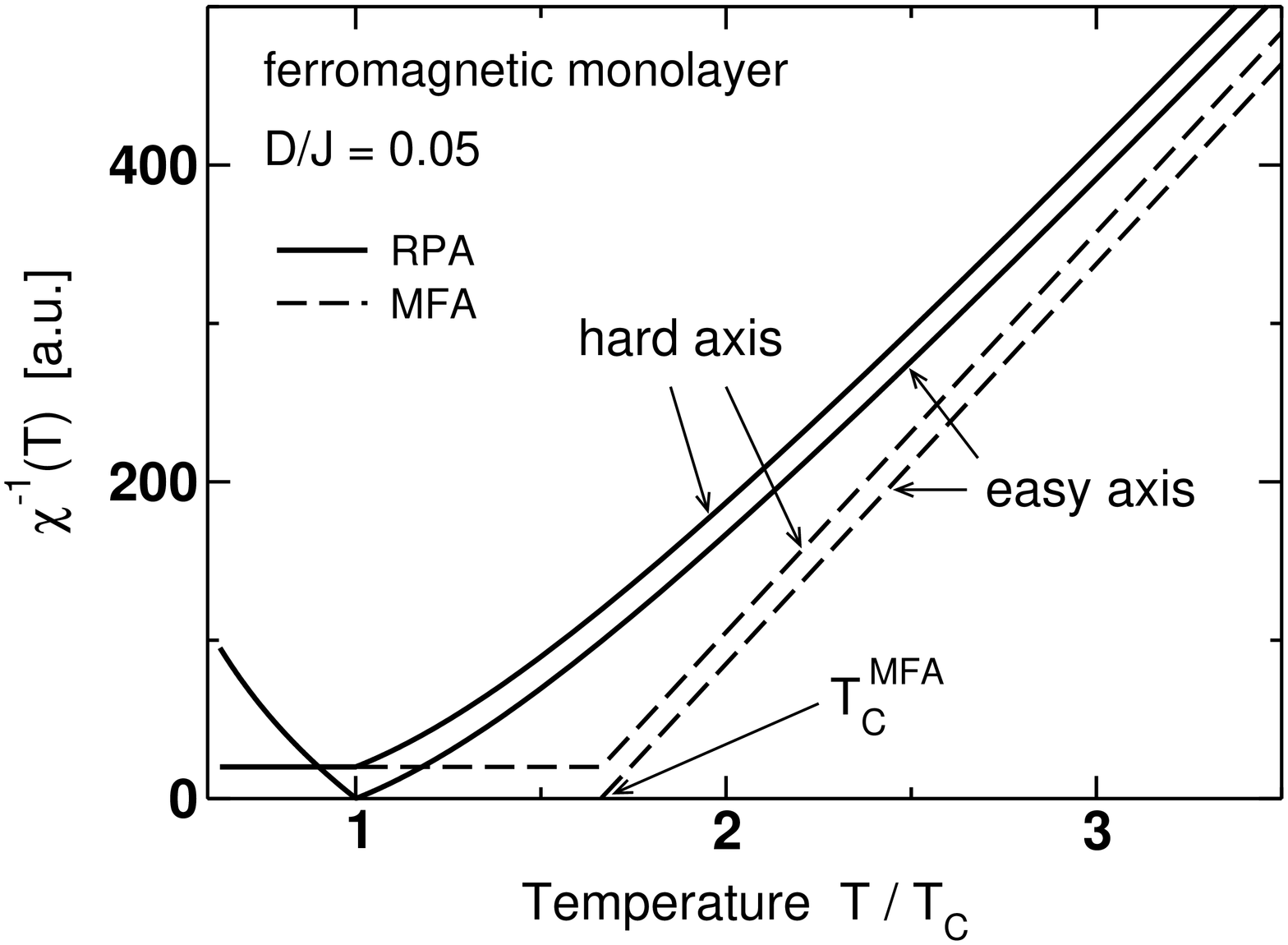}
\caption{
Calculated inverse parallel and transverse susceptibilities 
$\chi_{zz}^{-1}(T)$ 
and $\chi_{xx}^{-1}(T)$ along the easy and hard magnetic directions 
of a ferromagnetic (001) monolayer as function of temperature  
(full lines). The calculations have been performed with a Green's 
function approach (RPA), using Eqs.(\ref{e3a}) and (\ref{e3b}). 
For the exchange anisotropy we have assumed $D/J=0.05$, with $J$ the 
isotropic Heisenberg exchange between neighboring spins. 
The temperature is given in units of the Curie temperature 
$k_BT_C\equiv T_C^\mathrm{RPA}=0.63\,J$. For comparison the inverse 
susceptibilities as obtained from the mean field approximation (MFA) 
are shown (dashed lines), 
yielding the Curie temperature $k_BT_C^\mathrm{MFA}=1.05\,J$. Here 
$\chi_{zz}^{-1}(T)$ is depicted only for $T>T_C^\mathrm{MFA}$. 
}\end{figure} 

At first we have carried out corresponding calculations for the parallel 
susceptibility of a fcc (001) ferromagnetic film with two atomic 
layers using Eq.(\ref{e9}). The atomic magnetic moment 
$\mu_\mathrm{Co}=1.7\,\mu_B$ and the atomic volume 
$v_\mathrm{Co}=1.1\cdot10^{-29}\,\mathrm{m}^3$ appropriate for
bulk Co are assumed. The exchange coupling $J$ and the exchange 
anisotropy $D$ are obtained by fitting the results calculated with 
Eq.(\ref{e9}) to the measured $\chi_{zz}(T)$ in the temperature range 
between $T=T_C$ and $T=143{}^\circ$C. 
We obtain $J\,S(S+1)=44.1$~meV/bond, which is close to the Co bulk
value $J_\mathrm{Co}S(S+1)=39$~meV/bond. Furthermore, we get 
$D=7.0$~mK/atom for the exchange anisotropy, which corresponds to the 
single-ion in-plane uniaxial anisotropy 
$K_2=(1/2)(q_\|+q_\perp)\,D=28\,\mathrm{mK/atom}=2.4\,\mu\mathrm{eV/atom}$. 
From this value the anisotropy energy density is calculated to be 
$K'_2=K_2/v_\mathrm{Co}=3.3\cdot10^4\,\mathrm{J/m^3}$. 
This value should be compared with the one obtained from independent 
measurements of the effective anisotropy 
for the same system.\cite{WKO96} By means of a thermodynamic 
perturbation theory\cite{Cal66} we calculate the \textit{microscopic} 
anisotropy energy density at $T=0$ to be 
$K_2'=2\cdot10^4\,\mathrm{J/m^3}$ from the data of Ref.\cite{WKO96}, 
which is in reasonable agreement with the result based on the measured 
susceptibilities presented here. The resulting small value of 
$K'_2$ is comparable to the 6th-order anisotropy energy density 
$K_{66}^\mathrm{bulk}=1.2\cdot10^4\,\mathrm{J/m^3}$ in the 
hexagonal plane, and is about 20 times smaller than the 2nd-order uniaxial 
anisotropy energy density $K_2^\mathrm{bulk}=76.6\cdot10^4\,\mathrm{J/m^3}$ 
of bulk hcp Co.\cite{hcpCo} The calculated inverse parallel susceptibility 
$\chi_{zz}^{-1}(T)$ is depicted in Fig.2. A good 
agreement with experiment is obtained. We note that 
an increase of the anisotropy $D$ will result in a corresponding 
decrease of $\chi_{zz}(T)$ for $T>T_C$. 

We emphasize that the susceptibilities measured in the accessible
temperature range up to 155${}^\circ$C 
is still far below the linear (Curie-Weiss-) range. 
As mentioned in the Introduction, the paramagnetic Curie
temperature extrapolated from the inverse susceptibilities 
as calculated by the Green's function method is equal to the Curie
temperature $T_C^\mathrm{MFA}$ obtained from the mean field 
approximation.\cite{Tya67} For the parameters $J$ and $D$ 
as given above we obtain $T_C^\mathrm{MFA}=1091{}^\circ\mathrm{C}=1364$~K. 
Only for temperatures above $T_C^\mathrm{MFA}$ a Curie-Weiss 
behavior will emerge. The difference between $T_C$ and 
$T_C^\mathrm{MFA}$ and thus the range of the curved behavior of 
the inverse susceptibilities depicted in Fig.2 is determined to be 
very large for this ultrathin film. The reason is that the influence 
of collective magnetic fluctuations is much 
stronger in such two-dimensional systems as compared to bulk magnets. 

Finally, we like to demonstrate the behavior of $\chi_{zz}^{-1}(T)$ 
and $\chi_{xx}^{-1}(T)$ in a large temperature range above $T_C$.  
In Fig.3 the calculated inverse parallel and transverse suceptibilities 
for a \textit{single} square ferromagnetic layer are shown, using 
Eqs.(\ref{e3a}) and (\ref{e3b}). In order to reveal clearly the
difference between $\chi_{zz}^{-1}(T)$ and $\chi_{xx}^{-1}(T)$ a 
strong exchange anisotropy $D/J=0.05$ is assumed for these calculations. 
The temperature is given in units of $T_C=T_C^\mathrm{RPA}=0.63\,J$. 
For $T>T_C$ both susceptibilities differ by a temperature-independent shift, 
exhibiting the same curvature. The linear behavior of $\chi_{zz}^{-1}(T)$ 
and $\chi_{xx}^{-1}(T)$ is reached for elevated temperatures $T\gg T_C$,  
where the inverse susceptibilities approach the ones obtained from  
the MFA. Their extrapolations 
yield the characteristic temperatures $k_BT_z=q_\|\,(J+D)/4$ and 
$k_BT_x=q_\|\,J/4$, which differ by the anisotropy $q_\|\,D/4$. 
The larger value $T_z$ is the paramagnetic Curie temperature 
$T_C^\mathrm{para}=T_C^\mathrm{MFA}=1.05\,J$. We point out that the behavior 
of the inverse susceptibilites as calculated by the MFA is even 
qualitatively wrong, since it predicts a Curie-Weiss behavior for 
$\chi^{-1}(T)$ for all temperatures above the Curie temperature. 

In contrast, both inverse susceptibilities exhibit a considerably 
different behavior for $T\le T_C$. $\chi_{zz}^{-1}(T)$ vanishes at
$T=T_C$ for an infinitely small 
modulation amplitude, and increases strongly for $T<T_C$. On the other 
hand, $\chi_{xx}^{-1}(T)$ merely exhibits a cusp at $T_C$, 
and assumes the constant value $\chi_{xx}^{-1}(T)=q_\|\,D$ 
in the ferromagnetic phase. 
Thus, one could directly extract the anisotropy $D$  
by measuring $\chi_{xx}^{-1}(T_C)$, and determine $J$ from  $T_C$. 
This is an alternative treatment to the one as applied for the 2~ML 
case where both coupling constants are derived from the parallel 
susceptibility $\chi_{zz}(T)$ alone. However, one has to make sure
that secondary processes in the ordered phase are not effective. 
The apparent peak of the transverse susceptibility observed in 
Ref.\cite{BAP94} was attributed to vortex- and domain wall-excitations.
These result in a non-constant behavior of $\chi_{xx}^{-1}(T)$ for 
$T<T_C$, as is also seen in our experiments presented in Fig.2. 
For a quantitative comparison 
such domain processes or multi-domain states have to be considered 
as well. These complications are not expected to occur for $T>T_C$. 

\section{Discussion and Conclusion}
We have measured the parallel and transverse magnetic 
susceptibilities $\chi_{zz}(T)$ and $\chi_{xx}(T)$ of a 2.2~ML Co film 
grown on a vicinal Cu(1~1~17) surface. Corresponding 
calculations have been performed within an anisotropic Heisenberg 
model by applying a many-body Green's function method. 
We have demonstrated that the Curie temperature $T_C$ and the 
susceptibility in the paramagnetic regime ($T\ge T_C$) gives access to
the exchange interaction $J$ and the anisotropy $D$ of ultrathin films. 

Using these coupling parameters a quantitative agreement between 
experiment and theory is obtained at least in the limited temperature 
range accessible by our measurements. It is evident that this 
temperature range is far below the crossover to the Curie-Weiss 
behavior. Any extrapolation $\chi_{zz}^{-1}(T)\to0$ from the 
experimental data, assuming apparently linear parts 
in Fig.2, will yield a wrong value for the paramagnetic Curie 
temperature $T_C^\mathrm{para}=T_C^\mathrm{MFA}$. 
To reach the linear behavior of the susceptibilities the measurements 
have to be extended to temperatures around $1100{}^\circ$C, which is 
impossible for the Co/Cu thin film system due to its instability
against surface diffusion and alloying. 

The good agreement between theory and experiment justifies a 
methodological generalization for exploring magnetic properties 
by investigating the susceptibility in the paramagnetic regime.  
While the effect of the anisotropy in the paramagnetic regime 
for bulk systems is known for a long time,\cite{PT77,SuW70} 
for ultrathin ferromagnetic films improved theoretical 
approaches considering collective magnetic fluctuations have to 
be applied. A successful 
realization of corresponding experiments is challenging as the 
susceptibilities have to be measured in very small modulation fields
with very high accuracy. The measurements should be extended 
to as large temperatures as possible in order to allow for 
a comparison with theory over a wide temperature range. From 
such a comparison values for isotropic exchange interactions and 
anisotropies as present in the Heisenberg Hamiltonian Eq.(\ref{e1}) 
can be extracted. Note that these quantities refer to constant 
\textit{microscopic} parameters and not to \textit{effective} 
anisotropies measured at finite temperatures.\cite{Cal66} 

In the previous study on Fe/W(110) films a quantitative comparison 
between experiment and theory has
not been drawn.\cite{BAP94} Here  a different theoretical approach 
has been applied, namely a renormalization group treatment, allowing 
at the most only for a qualitative comparison. Thus the coupling 
constants for thin films cannot be determined by this method. It
should be mentioned that the Polyakov renormalization
scheme applied in Ref.\cite{BAP94} has been strongly criticized. Several
authors argue that this scheme might not be applicable for two-dimensional 
ferromagnets.\cite{LeG92}  

In future theoretical work we will explore the behavior of the magnetic 
susceptibility with increasing film thickness $L$. The range of the 
curved behavior of $\chi(T)$ for $T>T_C(L)$, which is very pronounced 
for ultrathin films as considered in the present study, is expected to 
reduce for thick films, approaching the one of the corresponding 
bulk system. With an improved theory for general spin quantum numbers 
$S$ the consideration of single-ion anisotropies becomes feasible. 
Such anisotropies are more appropriate for $3d$- transition metal magnets. 
A similar treatment for a \textit{perpendicular} anisotropy needs the
additional consideration of the magnetic dipole coupling. 
Anisotropies with a different symmetry, e.g.\ a \textit{quartic} in-plane
anisotropy, are also accessible within the scope of such a treatment,
resulting in a considerably different behavior of the susceptibilities 
with respect to the uniaxial case. 

Acknowledgedment: P.J.J.\ gratefully acknowledges the  invitation and 
the hospitality of the I.\  Institute for Theoretical Physics of  
the Hamburg University. 

\end{document}